\documentclass[aip,rsi,twocolumn,showpacs,groupedaddress,reprint,numerical]{revtex4-1}

\usepackage{amsmath,amssymb,graphicx}
\usepackage{dcolumn}
\usepackage{braket}
\usepackage{amssymb}
\usepackage{amsmath}
\usepackage{latexsym}
\usepackage{mathrsfs}
\usepackage[sans]{dsfont}
\usepackage{xcolor}
\usepackage{lipsum}
\usepackage{url}
\usepackage[caption=false]{subfig}
\usepackage{commath}
\usepackage{graphicx,bm}
\usepackage{verbatim}

\usepackage[colorlinks=true, linkcolor=blue, urlcolor=blue, citecolor=blue, anchorcolor=blue]{hyperref}

\begin{document}

\title{Injection locking of a low cost high power laser diode at 461~nm}

\author{C.J.H. Pagett}
\thanks{The first two authors contributed equally to this work.}
\author{P.H. Moriya\textcolor[rgb]{0,0,1}{\normalfont\textsuperscript{a)}}}
\email[E-mail me at:]{paulohisao@ifsc.usp.br}
\author{R. Celistrino Teixeira}
\author{R.F. Shiozaki}
\author{M. Hemmerling}
\author{Ph.W. Courteille}
\affiliation{Instituto de F{\'i}sica de S{\~a}o Carlos, Universidade de S{\~a}o Paulo, 13560-970 S{\~a}o Carlos, SP, Brazil.}

\begin{abstract}
Stable laser sources at 461~nm are important for optical cooling of strontium atoms. In most existing experiments this wavelength is obtained by frequency doubling infrared lasers, since blue laser diodes either have low power or large emission bandwidths. Here, we show that injecting less than 10~mW of monomode laser radiation into a blue multimode 500~mW high power laser diode is capable of slaving at least 50\% of the power to the desired frequency. We verify the emission bandwidth reduction by saturation spectroscopy on a strontium gas cell and by direct beating of the slave with the master laser. We also demonstrate that the laser can efficiently be used within the Zeeman slower for optical cooling of a strontium atomic beam.
\end{abstract}


\maketitle

\section{Introduction}
Motivated by the existence of narrow and ultra narrow transitions in strontium atoms the demonstration of optical cooling and trapping \cite{Kurosu90,Katori99} and recently the Bose-Einstein condensation \cite{Stellmer09,MartinezdeEscobar09} of strontium gases propelled this species among the hottest candidates for optical frequency standards \cite{Katori03,Nicholson15}. The strongest atomic resonance, $(5s^2)^1S_0$~-~$(5s5p)^1P_1$, which has a transition wavelength of 461~nm and a linewidth of 30.5~MHz \cite{Nagel05}, serves as a cooling transition in magneto-optical traps down to a few mK. As the saturation intensity of this transition is quite large, $I_{sat}=40~\text{mW/cm}^2$, relatively large laser power on the order of several 100~mW is required for efficient cooling in magneto-optical trap arrangements. Most laboratories working with strontium obtain the necessary power by frequency doubling an infrared laser operated at 922~nm. Although efficient, this option is rather cumbersome and expensive.

High gain and stable operation frequently are antagonistic requirements in laser diodes, especially in the wavelength range near 461~nm, where the panoply of blue laser diodes available on the market is sparse. A possible remedy is the use of the injection locking technique. Here, a high-power laser, called the slave laser, sees its noise level strongly reduced by injection of light from a low-noise low-power master laser into the slave laser's active zone. Provided that the frequencies of the master laser and the free-running slave laser are sufficiently close, the injection forces the slave laser to operate exactly on the injected frequency with little noise. The higher the injected power, the larger is the allowable frequency offset between the seed laser and the slave laser's resonance \cite{Siegman-86}. Close to quantum-limited intensity and phase noise can be achieved with this technique \cite{Pantell65,Tang67,Stover66,Buczek73,Winkelmann11,NewFocus}.

Recently, Y.~Shimada \textit{et al.} used expensive 100~mW single-mode laser diodes from Nichia (Nichia, NDB4216E-E), specified for a wavelength between 450~nm and 460~nm, to perform spectroscopy on a strontium hollow cathode \cite{Shimada13}. Two identical laser diodes were used in this experiment: A first laser diode was forced to operate near the Sr transition wavelength by optical feedback within an extended cavity setup (ECDL). When more than 20~mW of its power were injected into a second, free-running laser diode of the same type, the authors reached up to 110~mW of useful single-mode laser emission from the injected diode.

\begin{figure}
	\centerline{\includegraphics[width=8.5 truecm]{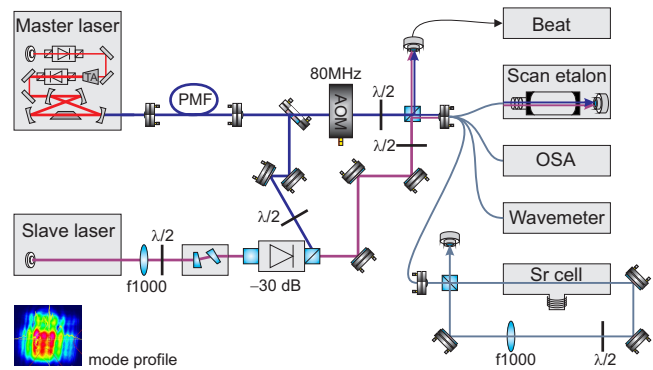}}
    \caption{(color online) Optical setup for injection locking. A single mode tunable master laser injects, through an optical isolator, the 
		slave laser diode. The emission spectra of both lasers are simultaneously monitored with a wavemeter, an optical spectrum analyzer (OSA), 
		and a scanning etalon. Also shown is the setup for beat frequency measurement and for saturation spectroscopy on the strong cooling 
		transition of strontium at 461~nm. (ND3: 3dB neutral density filter, PMF: polarization maintaining fiber, AOM: acousto-optic modulator)
		Also shown is a typical spatial mode profile of the slave laser observed with 200~mA laser current.}
    \label{fig:fig1}
\end{figure}

Cheap high power multi transverse mode laser diodes near 462~nm are nowadays sold for laser pointer and SciFi games applications \cite{Note1}. Here, we report on an injection locking experiment, where we harness the high power (nominally 1.4~W) multimode laser diode NDB7675 from Nichia with up to $\sim10~$mW of single-mode laser light. Another available option is the laser diode M462 which has 2~W nominal power\cite{Note2}. Both laser diodes, NDB7675 and M462, have similar spatial mode profile which suggests that the M462 laser diode is very likely to have similar properties as the NDB7675. The slave laser emits up to 430~mW of light power up to 50\% of which are slaved to the injected frequency. The limitation in the slave laser power comes from our stable current supply, which can afford a maximum current of 500~mA. The front window of the slave laser diode case is removed, in order to avoid competition between the injection locking and the optical feedback on the slave diode from the reflection on the front window. The main limitation arises from the multiple transverse modes emitted by the slave laser (see Fig.~\ref{fig:fig1}), which cannot perfectly be matched with the Gaussian mode profile of the master laser. Nevertheless, we find that the injected light efficiently narrows the slave laser emission bandwidth and even reduces the transverse mode spectrum.

We demonstrate the viability of the proposed scheme by a master-slave frequency beating measurement, by saturated absorption spectroscopy on a hot vapor strontium cell, and by laser-cooling of an atomic beam in a Zeeman slower. The master-slave locking scheme thus represents an efficient cheap and compact solution for a laser source driving the main strontium line and might reveal itself an asset in the future realization of transportable optical strontium clocks\cite{Poli14}. 

\section{Experimental set-up}
The experimental layout of the injection locking is shown in Fig.~\ref{fig:fig1}. The master laser, a Toptica TA-SHG pro, consists of an ECDL laser at 922~nm, amplified by a tapered amplifier and frequency-doubled to 461~nm. Part of its power is coupled to an optical fiber (PMF) delivering up to 10~mW power with a frequency bandwidth below 1~MHz. The main fraction of this light is injected through an optical isolator (Thorlabs, IO-3-460-HP) into the slave laser. The remaining power is frequency-shifted by 130~MHz with an acousto-optic modulator (AOM) and superposed with the light from the slave laser. The slave laser is the Nichia laser diode NDB7675. Its temperature is actively controlled with 10~mK precision. Both, the master and the slave laser frequencies are simultaneously monitored with a wavemeter (HighFinesse, WS6), an optical spectrum analyzer (HighFinesse, LSA), and a home-built scanning Fabry-P{\'e}rot etalon with a finesse of 50, as illustrated in Fig.~\ref{fig:fig1}. 
\begin{figure}
	\includegraphics[trim=50 350 80 30, clip,width=8.5 truecm]{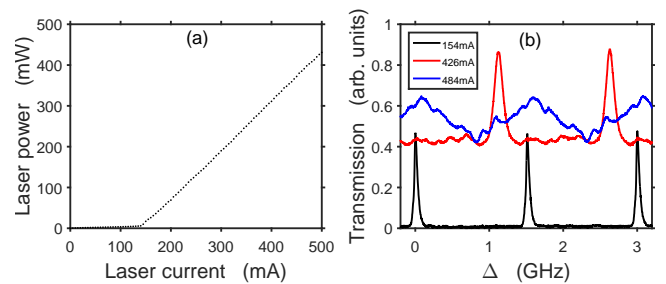}
    \caption{(color online) (a) Measured total emission power of the free-running laser diode as a function of laser current. 
		The lasing threshold is at a current of 140~mA, the slope efficiency is 1.2~W/A.
		(b) Emission spectrum of the free-running laser diode measured with a scanning Fabry-P{\'e}rot interferometer at low (black) and high (red and blue) laser currents.}
    \label{fig:fig2}
\end{figure}

The Nichia laser diode has a nominal power of 1.4~W with a several nanometer large spectrum centered at 462~nm and multiple transverse modes. Fig.~\ref{fig:fig2}(a) shows the measured laser power emitted by the free-running laser diode as a function of current.
Fig.~\ref{fig:fig2}(b) shows the emission spectrum recorded with a home-built scanning Fabry-P{\'e}rot interferometer with 1.5~GHz free spectral range and a finesse of 50. While at low (154~mA) current the spectrum is single mode, at higher current the spectrum becomes very broad, which is the cause for the large offset appearing in the spectrum of Fig.~\ref{fig:fig2}(b) for 426~mA current and the complete loss of the mode structure for 484~mA current.
\begin{figure}
	\includegraphics[trim=40 20 30 10, clip,width=9 truecm]{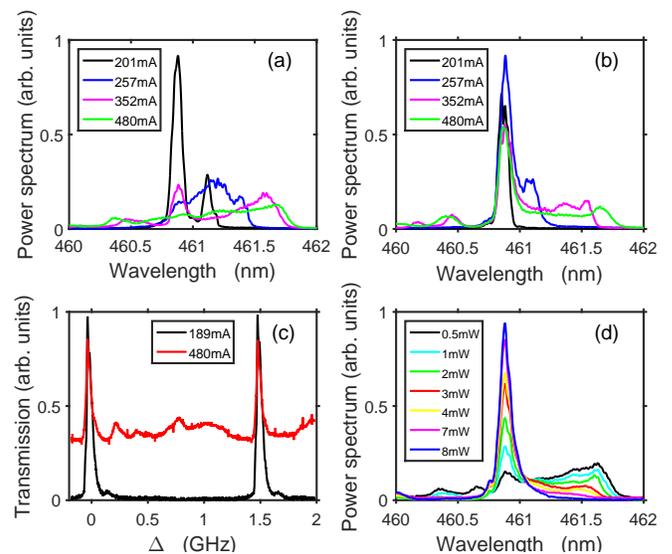}
    \caption{(color online) (a) Emission spectrum of the free-running laser diode measured with an optical spectrum analyzer at various laser currents. 
		The temperature of the laser diode was optimized for each current in order to maximize the power in the single emission peak close to the master frequency.
		(b) Same as (a) but with 0.5~mW injected light power from the master laser. 
		(c) Emission spectrum of the slave laser injected with 0.5~mW measured with a scanning Fabry-P{\'e}rot interferometer at low (black) and high (red) laser current.
		(d) Same as (a) but for a fixed laser current of 477~mA and varying the injection power.}
    \label{fig:fig3}
\end{figure}

\section{Characterization of injection locking}
In order to narrow the emission spectrum, we inject into the laser diode between 0.5 and 10~mW of light from the master laser. The spectra are recorded with an optical spectrum analyzer (High Finesse, LSA) with 6~GHz resolution and shown in Figs.~\ref{fig:fig3}(a-c). The spectra shown in Fig.~\ref{fig:fig3}(a) are obtained for various slave laser currents without injection. While at low current the power concentrates around a relatively narrow spectral region, at higher currents the spectra become much broader and almost continuous. Figs. \ref{fig:fig3}(b) are recorded with injection of 0.5~mW single mode radiation at 460.86~nm. At low currents the power spectrum appears to become single-mode near the master laser frequency. At higher currents, however, there is still a pedestal in the spectrum. This behavior of the slave laser is confirmed via the transmission spectrum of the scanning Fabry-P{\'e}rot etalon: Fig.~\ref{fig:fig3}(c) shows how, despite 0.5~mW injected power, the slave laser broadens its spectrum when its current is increased from 189~mA to 480~mW. The spectrum can further be purified by increasing the power of the injected light. Fig.~\ref{fig:fig3}(d) shows the narrowing of the spectra as the injection power is increased. At 10~mW of injection power no frequency components are discernible beyond the master laser frequency within the resolution of the optical spectrum analyzer.

\section{Measurements}
We demonstrate the performance of the injection locking system via three different measurements, 1.~a master-slave frequency beat, 2.~spectroscopy on a Sr cell, and 3.~application in laser-cooling.
\begin{figure}
	\includegraphics[width=8.5 truecm]{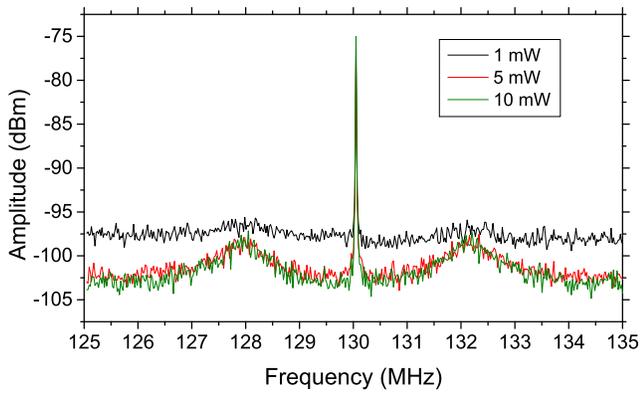}
   \caption{(color online) Beat frequency spectrum between the master laser frequency-shifted by an AOM at 130~MHz and the injected slave laser. 
		The line is stable and its resolution only limited by the resolution bandwidth (here 100~Hz) of the spectrum analyzer and the frequency-noise of the AOM. 
		The noise level is reduced with higher injection power at the exception of two small 1~MHz broad sidebands at $\pm2~$MHz to the left and right of the 
		beat frequency.}
    \label{fig:fig4}
\end{figure}

1.~We record the beat frequency spectrum between the slave laser and a reference beam obtained by frequency shifting the master laser by 130~MHz, as shown in Fig.~\ref{fig:fig1}. Within the 150~MHz bandwidth of the photodetector (Thorlabs, PDA10A) the spectrum (shown in Fig.~\ref{fig:fig4}) exhibits a single narrow line, whose width is only limited by the resolution of the spectrum analyzer. Interestingly, the noise floor drops with more efficient injection locking, leaving only two 1~MHz-wide sidebands, detuned 2~MHz from the beat frequency, which we suspect to be an effect of the bandwidth of the servo loop that stabilizes the frequency doubling cavity and corrects for fast noise on the laser current. This result shows that the injection of spectrally pure light is able to deplete the frequency components of the slave laser within a bandwidth of at least 150~MHz. The spectral purity of the beat spectrum in a large frequency band is encouraging for applications of the slave laser in high-resolution laser spectroscopy and narrow atomic transitions and optical cooling of atomic gases.

2.~These findings do, however, not exclude the possible presence of frequency components beyond the 150~MHz bandwidth. Such components should, in principle, appear in the spectra taken with the Fabry-P{\'e}rot interferometer [see Fig.~2(b)] or the optical spectrum analyzer [see Fig.~3(a-c)]. However, the sensitivity of the interferometer is too low and the resolution of the optical spectrum analyzer too broad.

In order to obtain information on the slave laser emission spectrum beyond the 150~MHz bandwidth, we performed saturation spectroscopy on a 15~cm long strontium gas cell heated to $530^\circ$C. The temperature, and hence the strontium partial pressure, are adjusted such as to yield a measured optical density on the order of 1. The cell also contained an argon buffer gas at pressure 1~mbar \cite{Note3}. As shown in Fig.~\ref{fig:fig1}, the saturated absorption signal is obtained by passing through the Sr cell a saturation beam (8~mW corresponding to 20\% of the saturation intensity) and a counter-propagating probe beam obtained by retro-reflecting an attenuated (6~dB) fraction of the saturation beam.
\begin{figure}
	\includegraphics[width=4.25 truecm]{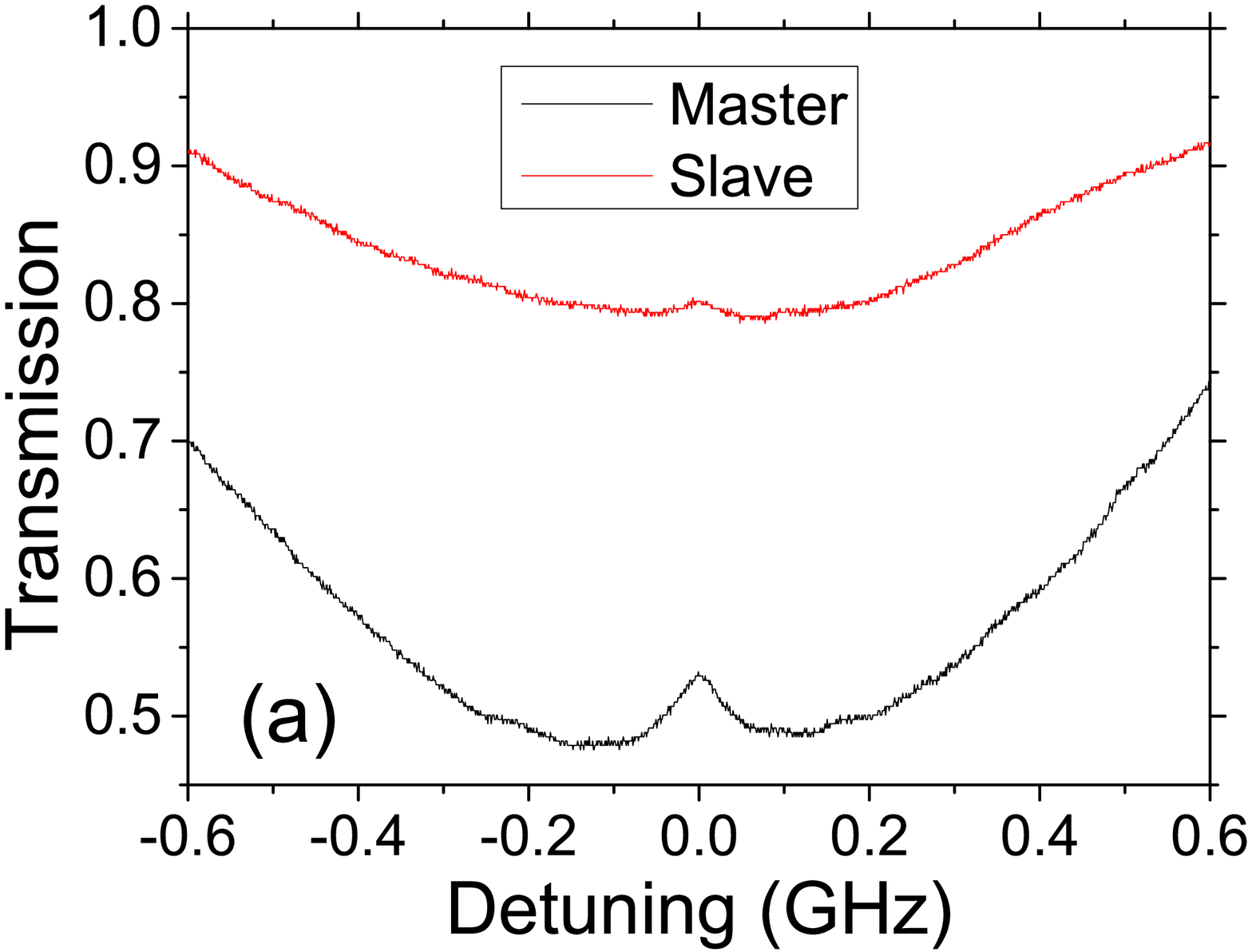}
	\hspace*{-0.25 truecm}
	\includegraphics[width=4.3 truecm]{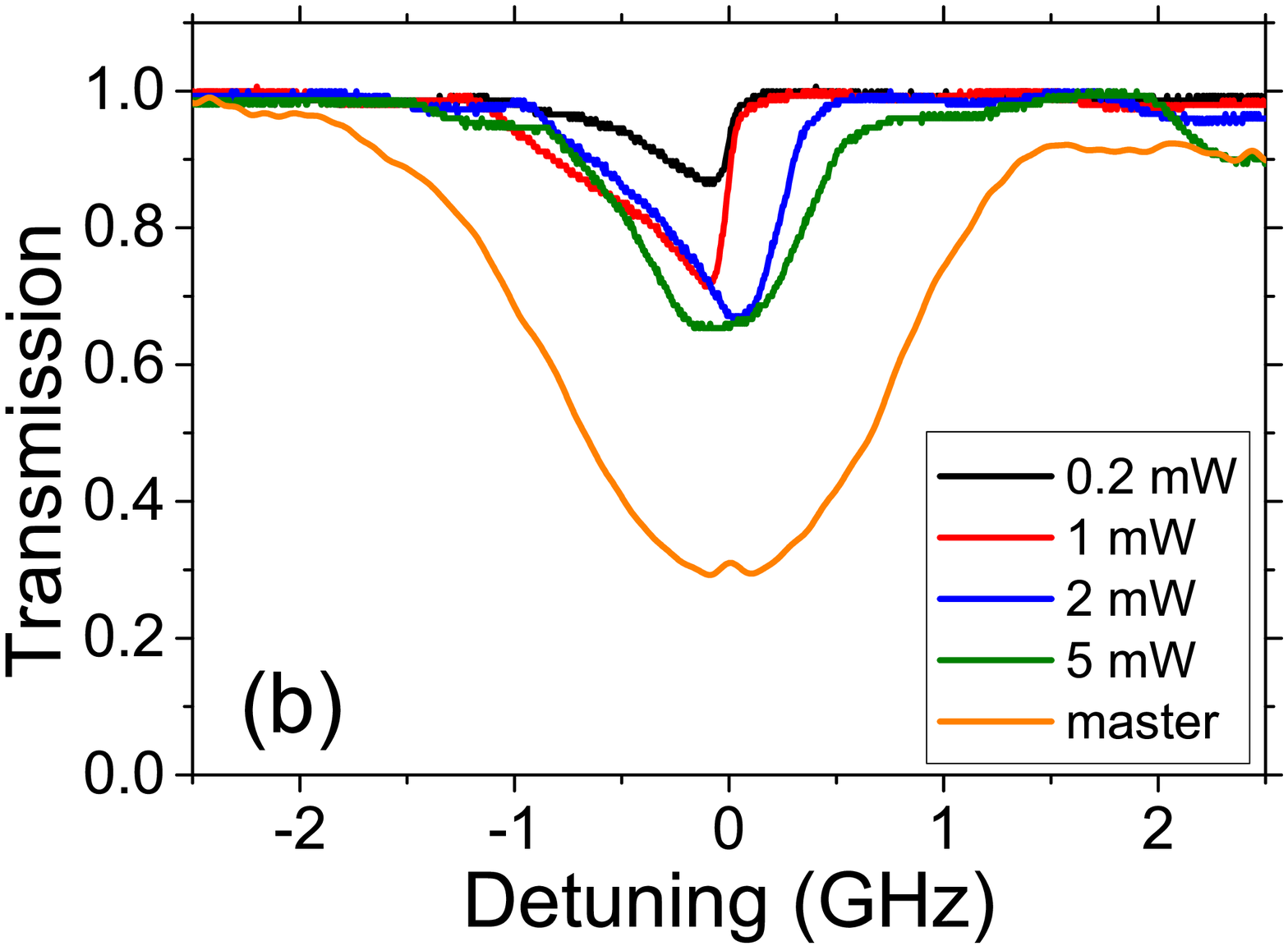}
    \caption{(color online) (a) Saturated absorption spectra recorded with the master laser (lower black curve) and with the 
		injection-locked slave laser (upper red curve). The injection power was 8.6~mW and the slave laser current 480~mA.
		Both laser beams had 8~mW power (20\% of the saturation intensity) at the entrance of the cell.
		(b) Single-pass absorption spectrum of the master laser (orange) and the slave laser (other lines) for various injection powers versus the master laser detuning $\Delta$ to the $^1S_0$~-~$^1P_1$ frequency transition, $\nu{_{Sr}}=650.503225$ THz.}
    \label{fig:fig5}
\end{figure}

The $^1S_0$~-~$^1P_1$ transition of $^{88}$Sr has a natural linewidth of 30.5~MHz. The master laser was ramped across the 2~GHz large Doppler profile of this transition. Fig.~\ref{fig:fig5}(a) shows two spectra: The black (lower) curve is taken with the master laser itself, while the red (upper) curve is taken with the slave laser. For all spectra recorded with the slave laser, we always scan the master laser's frequency. The injection locking is optimized at the center of the scan. Clearly visible is in both cases the Doppler-free Lamb dip. However, the spectrum taken with the slave laser exhibits a large offset caused by a constant large background of transmitted light.

An explanation could be the existence of frequency components in the slave laser spectrum, which are not resonant with the Sr line and, hence, not absorbed. To study this further, we show in Fig.~\ref{fig:fig5}(b) absorption spectra of the slave laser recorded at various injection laser powers. As seen in Fig.~\ref{fig:fig3}(c), at low injection laser power the slave laser emission spectrum is so broad that most of the laser power is transmitted. As the injection power is increased, the emission concentrates more and more near the Sr resonance frequency, so that the more power is absorbed. Moreover, the spectral narrowing of the absorption spectrum points towards an efficient narrowing of the slave laser emission spectrum. The total absorption ($1$ minus the transmission) shown at the center of the slave laser spectrum injected with $5~$mW of master light, as shown on Fig. \ref{fig:fig5}(b), is $0.35$. The total absorption at the center of the master laser spectrum is $0.71$. From that, we conclude that 50\% of the slave power was efficiently emitted within the desired frequency, while the remaining 50\% is out of resonance with the Strontium atoms and appears as a constant background for our absorption spectra. Note that the widths of the transmission profiles of Fig.~\ref{fig:fig5}(b) created by the slave laser are smaller than the width of the transmission profile for the master laser. This gives an estimation of the robustness of the injection locking when we scan the master frequency. Indeed, at about 1~GHz from the optimized injection locking frequency, no more light is absorbed from the slave laser, while the master still shows some absorption, indicating a very poor locking of the frequency of the slave.

3.~Up to this point we demonstrated spectral purity with a 150~MHz bandwidth and the existence of frequency components beyond 2~GHz. When used for spectroscopy at atomic resonances, such as Sr, these frequency components are sufficiently far away not to disturb the atoms (the atoms are basically transparent), so that it is likely that the setup can be used, e.g. for laser-cooling purposes. Our lab disposes of an experimental setup for the preparation of cold atomic $^{88}$Sr in a magneto-optical trap (MOT) loaded from a Zeeman slower: A chunk of strontium is heated to $580^\circ$C in an oven connected, via a two-dimensional array of microtubes, with a 35~cm long vacuum tube hosting a Zeeman slower in spin-flip configuration \cite{Courtillot03}, where it generates a collimated atomic beam. The Zeeman slower is operated with laser light derived from the master laser tuned 530~MHz below the atomic resonance frequency and injected in counter-propagating direction to the atomic beam. The Zeeman slower tube ends on an ultrahigh vacuum chamber, where the atoms are captured by the MOT and cooled to about 5~mK temperature. The fluorescence recorded from the MOT is a measure for the number of trapped atoms and, hence, of the efficiency of the Zeeman slower.
\begin{figure}
	\includegraphics[width=4.1 truecm]{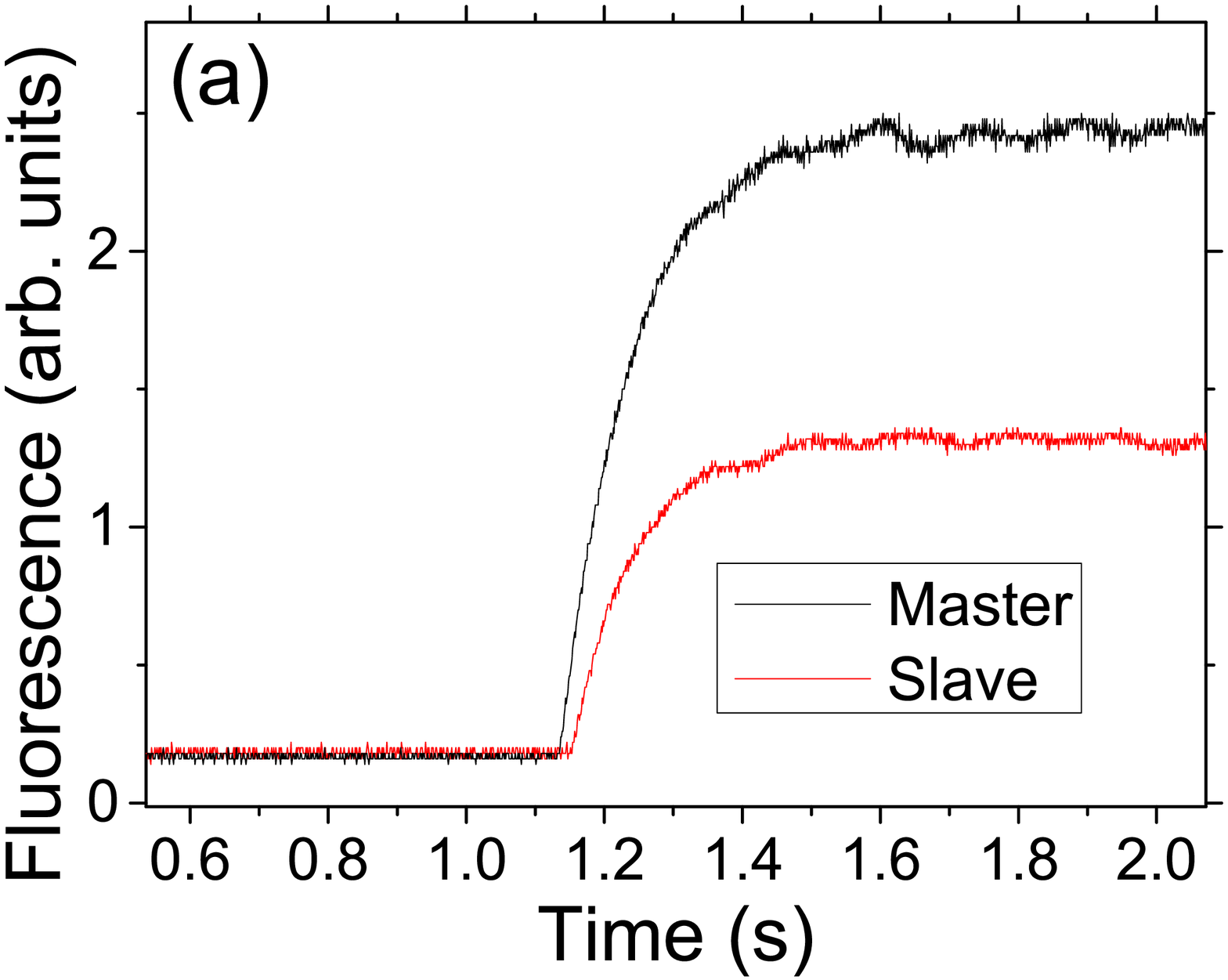}
	\hspace*{-0.1 truecm}
	\includegraphics[width=4.3 truecm]{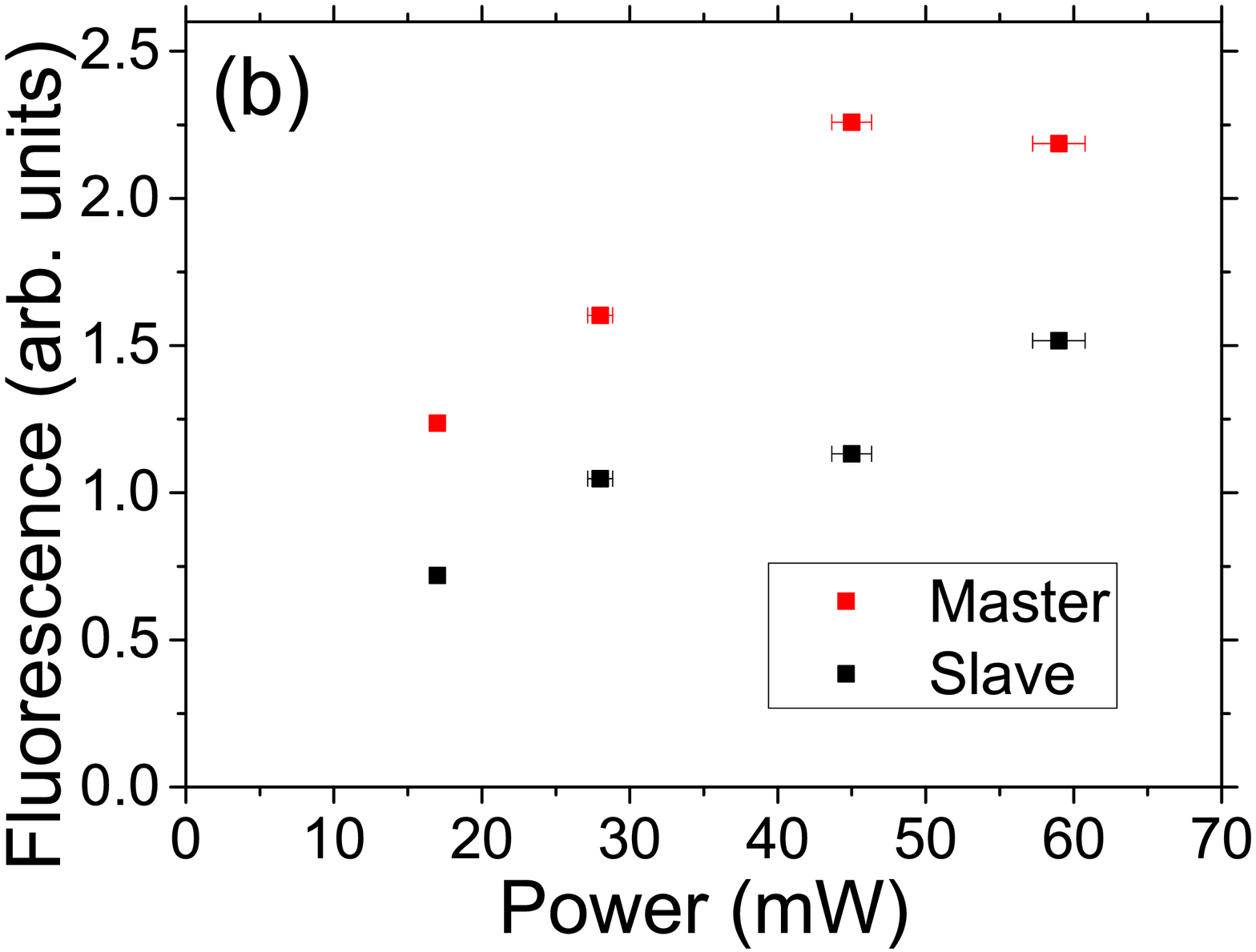}
    \caption{(color online) (a) MOT fluorescence recorded while switching the light operating the Zeeman slower: 
		At the time 1.1~s the master (black) or the slave (red) laser is suddenly irradiated into the Zeeman slower, 
		thus loading the MOT with Sr atoms. In both cases, the master and the slave laser had the same power of 45~mW.
		(b) Fluorescence measured after the MOT has been fully loaded for various lasers powers irradiated into the 
		Zeeman slower using the master (black) or the slave (red) laser.}
    \label{fig:fig6}
\end{figure}

In order to demonstrate that the injected slave laser light can be used for laser-cooling purposes, instead of sending the Zeeman slower light beam to the main experimental setup, we use it as the master laser for the injection locking scheme, and then send the slave laser light to the Zeeman slower. To perform these measurements we have injected the slave laser diode with 8~mW. Before the slave laser beam shines the Zeeman slower, it passes through an iris to improve its shape which results in a operational power of 60~mW. Fig. \ref{fig:fig6}(a) shows the resonance fluorescence of the magneto-optical trap when we turn on the light operating the Zeeman slower for both cases, i.e., the light is taken directly from the master (black curve) or indirectly from the injected slave laser. The MOT is operated by light directly derived from the master laser.

The efficiency of the Zeeman slower is smaller for the slave as compared to the master for the same light intensity, because part of the slave power is in far away frequencies that do not contribute to the deceleration effect. This decrease in efficiency is compatible with the previous measured 50\% efficiency of the injection locking. Nevertheless, the nearly linear behavior of the MOT fluorescence as a function of the slave laser intensity, as shown in Fig. \ref{fig:fig6}(b), strongly suggests that the Zeeman slower efficiency would still benefit from an increase of the slave laser power. Our setup prevented the observation of this improvement, because our stable current supply was limited to 500~mA, while the slave diode can afford a forward current up to 1.7~A.

\section{Conclusion}
In conclusion, we found that injection-locked multi-transverse mode laser diodes can profitably be used in high-resolution spectroscopy. Operated at high laser powers up to 500~mW their transverse mode profile becomes increasingly broad, which hampers a perfect mode-matching of the injection beam. Nevertheless, at sufficiently high injection power of several mW, a frequency band of at least 150~MHz width is cleared from noise. As we have shown, this is sufficient for applying it as part of the laser cooling process of strontium atoms, more precisely into a Zeeman Slower.

In our setup the master laser power available for injection was limited to 10~mW, but we expect that even higher injection power will improve the slave laser stability, as has been shown by Y.~Shimada \textit{et al.} \cite{Shimada13} using single-mode laser diodes. In future work, the TA SHG pro master laser will be replaced by a home-built ECDL laser in Littrow configuration operating with a Nichia NDB4216E-E laser diode delivering up to 40~mW of laser power, as it has been done by Shimada \textit{et al.}\cite{Shimada13}, resulting in a fully low cost system.

\bigskip\noindent\textbf{\emph{Acknowledgements.\---~}} This work was funded by the Funda\c{c}{\~a}o de Amparo {\`a} Pesquisa do Estado de S{\~a}o Paulo (FAPESP). We thank Luiz G. Marcassa for lending us the HighFinesse spectrometer.

\end{document}